\documentclass[aps,nofootinbib,preprintnumbers]{revtex4}
\usepackage{eurosym}
\usepackage[utf8]{inputenc}
\usepackage[english]{babel}
\usepackage{lipsum}
\usepackage{amsmath, amssymb, amsfonts}
\usepackage{bm}
\usepackage{slashed}
\usepackage{bbm}
\usepackage{xcolor}
\usepackage{graphicx}
\usepackage{booktabs}
\usepackage{multirow}
\usepackage{float}
\usepackage{placeins}
\usepackage{verbatim}
\usepackage{todonotes}
\usepackage[colorlinks=true,urlcolor=red,citecolor=red]{hyperref}

\newenvironment{Eqnarray}{\arraycolsep 0.14em\begin{eqnarray}}{\end{eqnarray}}
\newcommand{\ba}{\begin{Eqnarray}}
\newcommand{\ea}{\end{Eqnarray}}
\newcommand{\be}{\begin{equation}}
\newcommand{\ee}{\end{equation}}
\newcommand{\bal}{\begin{aligned}}
\newcommand{\eal}{\end{aligned}}
\newcommand{\bea}{\begin{eqnarray}}
\newcommand{\eea}{\end{eqnarray}}
\newcommand{\ben}{\begin{enumerate}}
\newcommand{\een}{\end{enumerate}}
\newcommand{\bit}{\begin{itemize}}
\newcommand{\eit}{\end{itemize}}
\newcommand{\bde}{\begin{widetext}}
\newcommand{\ede}{\end{widetext}}

\renewcommand{\[}{\left[}

\def\lsim{\mathrel{\rlap{\lower4pt\hbox{\hskip1pt$\sim$}}
\raise1pt\hbox{$<$}}}
\def\gsim{\mathrel{\rlap{\lower4pt\hbox{\hskip1pt$\sim$}}
\raise1pt\hbox{$>$}}}

\def\3211{$\mathrm{SU(3) \otimes SU(2)_L \otimes U(1)_R \otimes U(1)_{B-L}}$ }
\def\321{$\mathrm{SU(3) \otimes SU(2) \otimes U(1)}$ }
\def\422{$\mathrm{SU(4) \otimes SU(2) \otimes SU(2)_R}$ }

\definecolor{bostonuniversityred}{rgb}{0.8, 0.0, 0.0}

\input{tcilatex}
\begin{document}

\title{Cobimaximal mixing pattern from a $\Delta \left( 27\right) $
inverse seesaw model.}
\author{A. E. C\'arcamo Hern\'andez$^{a,b,c}$}
\email{antonio.carcamo@usm.cl}
\affiliation{$^{{a}}$Universidad T\'ecnica Federico Santa Mar\'{\i}a, Casilla 110-V,
Valpara\'{\i}so, Chile\\
$^{{b}}$Centro Cient\'{\i}fico-Tecnol\'ogico de Valpara\'{\i}so, Casilla
110-V, Valpara\'{\i}so, Chile\\
$^{{c}}$Millennium Institute for Subatomic Physics at the High-Energy
Frontier, SAPHIR, Calle Fern\'andez Concha No 700, Santiago, Chile}
\author{Ivo de Medeiros Varzielas}
\email{ivo.de@udo.edu}
\affiliation{CFTP, Departamento de F\'{\i}sica, Instituto Superior T\'{e}cnico,
Universidade de Lisboa, Avenida Rovisco Pais 1, 1049 Lisboa, Portugal}
\author{Nicol\'{a}s A. P\'{e}rez-Julve}
\email{nicolasperezjulve@gmail.com}
\affiliation{Departamento de F\'isica, Universidad T\'{e}cnica Federico Santa Mar\'{\i}a}
\affiliation{Millennium Institute for Subatomic Physics at high energy frontier - SAPHIR,
Fernandez Concha 700, Santiago, Chile}
\date{\today }

\begin{abstract}
We present an inverse 
seesaw model based on the $\Delta(27)$ symmetry and Abelian discrete symmetries, which account for the mass hierarchies and, through a specific pattern of symmetry breaking,  leads to viable leptonic mixing angles according to the cobimaximal mixing pattern. In the model, leptogenesis successfully accounts for the observed  baryon asymmetry of the Universe for a large range of the parameter space, only for the scenario of normal neutrino mass hierarchy.
\end{abstract}

\maketitle



\section{Introduction}

\label{intro} The discovery of the Higgs boson at the LHC has provided
additional confirmation of the Standard Model (SM). Nevertheless, certain
phenomena such as neutrino oscillations require physics beyond the SM.
Within the SM, fermion masses and mixing angles originate from Yukawa
couplings to the Higgs field. These couplings are numerous, hierarchical,
and exhibit markedly different patterns between quarks and leptons,
presenting a flavour problem. Whereas in the quark sector the mixing angles
are small, in the lepton sector two of the mixing angles are large and one
is small. A potential resolution to this flavor problem is to consider
extensions of the SM with enlarged particle content as well as
extended symmetries, like for example family symmetries, often combined with
a seesaw mechanism to generate the tiny active neutrino masses. Such
symmetries, when broken in a specific pattern, can explain the structure of
the Yukawa couplings. One well-studied family symmetry is $\Delta (27)$,
which offers several advantages: it is relatively small, includes triplet
and anti-triplet representations, and naturally accommodates the observed
leptonic mixing in many models. It also exhibits interesting connections
with CP symmetry \cite{Branco:1983tn, deMedeirosVarzielas:2006fc, Ma:2006ip,
Ma:2007wu, Bazzocchi:2009qg, deMedeirosVarzielas:2011zw, Varzielas:2012nn,
Bhattacharyya:2012pi, Ferreira:2012ri, Ma:2013xqa, Nishi:2013jqa,
Varzielas:2013sla, Aranda:2013gga, Varzielas:2013eta, Harrison:2014jqa,
Ma:2014eka, Abbas:2014ewa, Abbas:2015zna, Varzielas:2015aua,
Bjorkeroth:2015uou, Chen:2015jta, Vien:2016tmh, Hernandez:2016eod,
Bjorkeroth:2016lzs, CarcamoHernandez:2017owh, deMedeirosVarzielas:2017sdv,
Bernal:2017xat, CarcamoHernandez:2018iel, deMedeirosVarzielas:2018vab,
CarcamoHernandez:2018hst, CarcamoHernandez:2018djj, Ma:2019iwj,
Bjorkeroth:2019csz, CarcamoHernandez:2020udg, CarcamoHernandez:2024vcr}. In
this work, we propose a minimal model based on $\Delta (27)$ family symmetry
and the inverse seesaw mechanism responsible for the generation of the tiny
active neutrino masses. To achieve a viable model, $\Delta (27)$ must be
broken by several fields, a common feature in models with discrete triplet
representations \cite{Bhattacharyya:2012pi, Varzielas:2013sla,
Varzielas:2013eta, CarcamoHernandez:2024vcr}. Our focus here is exclusively
on leptons; the quark sector can be treated as in the SM. The layout of the paper goes as
follows. In section \ref{sec:model} we describe the model with its
symmetries, particle content and Yukawa interactions. The implications of
the model in lepton masses and mixings are discussed in section \ref%
{leptonmixings}. In section \ref{lepto} we analyze its consequences on the
baryon asymmetry of the Universe (BAU) through leptogenesis. Our conclusions
are stated in section \ref{conclusions}.

\section{The model}

\label{sec:model} We consider a supersymmetric model where the tiny masses
of the active neutrinos are generated from an inverse seesaw mechanism and
the leptonic mixing features the cobimaximal mixing pattern \cite{Fukuura:1999ze,Miura:2000sx,Ma:2002ce,Grimus:2003yn,Chen:2014wxa,Ma:2015fpa,Joshipura:2015dsa,Li:2015rtz,He:2015xha,Chen:2015siy,Ma:2016nkf,Damanik:2017jar,Ma:2017trv,Grimus:2017itg,CarcamoHernandez:2017owh,CarcamoHernandez:2018hst,Ma:2019iwj,Hernandez:2021kju,Rivera-Agudelo:2022qpa,CarcamoHernandez:2024ycd, Rivera-Agudelo:2024vdn}. In order to
implement the cobimaximal mixing pattern of the leptonic mixing we enlarge
the SM gauge symmetry by the inclusion of the $\Delta\left(27\right)$ family
symmetry as well as the auxiliary cyclic symmetries $Z_2$, $Z_3$ and $Z_9$.
In the supersymmetric framework proposed here, the MSSM scalar sector
composed of the two $SU(2)_L$ doublet scalars $H_u$ and $H_d$ is augmented
by a set of additional gauge-singlet scalar fields $\sigma$, $\phi$, $\rho$
and $\eta$, while the fermionic content is extended to include six heavy
right-handed Majorana neutrinos $\nu_{iR}$ and $N_{iR}$ ($i=1,2,3$), which
are singlets under the SM gauge group. The presence of these heavy states
enables the implementation of the inverse seesaw mechanism, which naturally
accounts for the small masses of the light active neutrinos. The model
particle content and the field assignments under the $\Delta(27)\times
Z_2\times Z_3 \times Z_9$ discrete group is displayed in Table \ref{model}.
In Table \ref{model} the $\Delta(27)$ representations are expressed by
numbers in boldface, whereas the charges of the fields under the different
cyclic symmetries are given in additive notation. The inclusion of the $%
\Delta(27)$ family symmetry in our model is necessary to get a predictive
model for the leptonic mixings, which features the cobimaximal mixing
pattern. The $Z_3$ symmetry separates the $\Delta(27)$ scalar triplet $\eta$
participating in the charged lepton Yukawa interactions and in the Yukawa
terms given subleading contributions to the Dirac neutrino submatrix from
the one ($\phi$) appearing in the Yukawa terms responsible for the
generation of the lepton number violating (LNV) Majorana mass parameter $\mu$
of the inverse seesaw. The $Z_9$ symmetry is responsible for generating the
SM charged lepton mass pattern. As follows from the properties of the cyclic
groups, let us note that the $Z_9$ symmetry is the smallest cyclic symmetry
in order to get the thirteen dimensional Yukawa operator $\left( \overline{l}%
_{L}H_{d}\eta \right) _{\mathbf{1}_{0,0}}l_{1R}\frac{\sigma ^{8}}{\Lambda^{9}%
}$ from the $\frac{\sigma^{8}}{\Lambda^{8}}$ insertion on the $\left( 
\overline{l}_{L}H_{d}\eta \right) _{\mathbf{1}_{0,0}}l_{1R}$ operator which
is responsible for the $\lambda^{9}$ suppression required to naturally yield a small electron mass. The $Z_2$
symmetry prevents the five dimensional Yukawa term $\left( \overline{l}%
_{L}H_{d}\eta \right)_{\mathbf{1}_{0,0}}l_{1R}\frac{\rho}{\Lambda}$, whose
inclusion in the model would require to set an unnatural small value for its
corresponding Yukawa coupling, in order to successfully reproduce the
experimental value of the electron mass (we consider $\lambda=0.225$). To achieve the cobimaximal mixing
pattern for the leptonic mixings we consider the following vacuum
expectation value (VEV) configurations of the $\Delta(27)$ triplet scalars $%
\phi$ and $\rho$: 
\begin{equation}
\left\langle \eta \right\rangle =v_{\eta }\left( 1,0,0\right) ,\hspace{1cm}
\left\langle \phi \right\rangle =v_{\phi }\left( 1,\omega ,\omega
^{2}\right) ,\hspace{1cm}\omega =\exp \left( \frac{2\pi i}{3}\right) , 
\notag
\end{equation}
To produce the observed SM charged fermion mass hierarchy via spontaneous $%
Z_9$ symmetry breaking, we therefore define the VEVs for the relevant gauge singlet scalars in the following way: 
\begin{equation}
v_{\rho }\sim v_{\sigma }\sim v_{\phi }\sim v_{\eta }\sim \lambda \Lambda ,
\label{VEVhierarchy}
\end{equation}%
where $\lambda =\sin \theta _{13}$ is the reactor mixing parameter and $%
\Lambda $ is the the scale of the UV completion of the model.

The supersymmetric model under consideration features a two-step spontaneous
symmetry breaking of the full symmetry group $\mathcal{G}$: 
\begin{eqnarray}
&&\mathcal{G}=SU(3)_{C}\times SU(2)_{L}\times U(1)_{Y}\times\Delta(27)\times
Z_2\times Z_3 \times Z_9  \notag \\
&&\hspace{35mm}\Downarrow \Lambda _{int}  \notag \\[3mm]
&&\hspace{15mm}SU(3)_{C}\times SU(2)_{L}\times U(1)_{Y}  \notag \\[3mm]
&&\hspace{35mm}\Downarrow v  \notag \\[3mm]
&&\hspace{23mm}SU(3)_{C}\otimes U(1)_{em}  \label{Group}
\end{eqnarray}
This spontaneous symmetry breaking is realized at high energies in a sequential pattern. The discrete subgroup $\Delta(27)\times
Z_2\times Z_3 \times Z_9$ is broken at an intermediate scale $\Lambda_{int}$%
, which is assumed to be much larger than the electroweak scale $v = 246\;%
\text{GeV}$. The latter triggers the standard breaking of the electroweak
symmetry down to electromagnetism. 
\begin{table}[tbp]
\begin{eqnarray*}
\begin{array}{|c|c|c|c|c|c|c|c|c|c|c|c|c|}
\hline
& l_{L} & l_{1R} & l_{2R} & l_{3R} & \nu _{R} & N_{R} & H_{u} & H_{d} & 
\sigma & \eta & \phi & \rho \\ \hline
\Delta \left( 27\right) & \mathbf{3} & \mathbf{1}_{0,0} & \mathbf{1}_{0,1} & 
\mathbf{1}_{0,2} & \mathbf{3} & \overline{\mathbf{3}} & \mathbf{1}_{0,0} & 
\mathbf{1}_{0,0} & \mathbf{1}_{0,0} & \mathbf{3} & \mathbf{3} & \mathbf{1}%
_{1,0} \\ \hline
Z_{2} & 0 & 0 & 0 & 0 & 0 & 0 & 0 & 0 & 0 & 0 & 0 & 1 \\ \hline
Z_{3} & 0 & 1 & 1 & 1 & 0 & 0 & 0 & 0 & 0 & -1 & 0 & 0 \\ \hline
Z_{9} & 0 & 8 & 4 & 2 & 0 & 0 & 0 & 0 & -1 & 0 & 5 & 1 \\ \hline
\end{array}%
\end{eqnarray*}
\vspace{-0.4cm}
\caption{The transformation properties of the scalar and fermionic fields
under the $\Delta (27)\times Z_{2}\times Z_{3}\times Z_{9}$ discrete group.}
\label{model}
\end{table}
With the model particle content and symmetries displayed in Table \ref{model}%
, we can build the following Yukawa terms for the charged lepton and
neutrino sectors: 
\begin{equation}
-\mathcal{L}_{Y}^{\left( l\right) }=y_{1}^{\left( l\right) }\left( \overline{%
l}_{L}H_{d}\eta \right) _{\mathbf{1}_{0,0}}l_{1R}\frac{\sigma ^{8}}{\Lambda
^{9}}+y_{2}^{\left( l\right) }\left( \overline{l}_{L}H_{d}\eta \right) _{%
\mathbf{1}_{0,2}}l_{2R}\frac{\sigma ^{4}}{\Lambda ^{5}}+y_{3}^{\left(
l\right) }\left( \overline{l}_{L}H_{d}\eta \right) _{\mathbf{1}_{0,1}}l_{3R}%
\frac{\sigma ^{2}}{\Lambda ^{3}}+H.c.  \label{Lyl}
\end{equation}
\begin{eqnarray}
-\mathcal{L}_{Y}^{\left( \nu \right) } &=&y_{\nu }\left( \overline{l}%
_{L}H_{u}\nu _{R}\right) _{\mathbf{1}_{0,0}}+\sum_{k=0}^{2}\sum_{l=0}^{2}x_{%
\nu }^{\left( k,l\right) }\left( \overline{l}_{L}H_{u}\nu _{R}\right) _{%
\mathbf{1}_{k,l}}\frac{\left[ \left( \eta \eta \right) _{\overline{\mathbf{3}%
}_{S_{1}}}\eta \right] _{\mathbf{1}_{3-k,3-l}}}{\Lambda ^{3}}+z_{\nu }\left( 
\overline{l}_{L}H_{u}\nu _{R}\right) _{\mathbf{1}_{1,0}}\frac{\sigma
^{2}\rho ^{2}}{\Lambda ^{4}}  \notag \\
&&+m\left( \overline{\nu }_{R}N_{R}^{C}\right) _{\mathbf{1}%
_{0,0}}+\sum_{k=0}^{2}\sum_{l=0}^{2}y_{\nu N}^{\left( k,l\right) }\left( 
\overline{\nu }_{R}N_{R}^{C}\right) _{\mathbf{1}_{k,l}}\frac{\left[ \left(
\eta \eta \right) _{\overline{\mathbf{3}}_{S_{1}}}\eta \right] _{\mathbf{1}%
_{3-k,3-l}}}{\Lambda ^{3}}+y_{1}^{\left( N\right) }\left( \overline{N}%
_{R}N_{R}^{C}\right) _{\overline{\mathbf{3}}_{S_{1}}}\phi \frac{\sigma ^{5}}{%
\Lambda ^{5}}  \label{Lynu} \\
&&+y_{2}^{\left( N\right) }\left( \overline{N}_{R}N_{R}^{C}\right) _{%
\overline{\mathbf{3}}_{S_{2}}}\phi \frac{\sigma ^{5}}{\Lambda ^{5}}%
+y_{3}^{\left( N\right) }\left( \overline{N}_{R}N_{R}^{C}\right) _{\overline{%
\mathbf{3}}_{S_{1}}}\phi \frac{\rho ^{4}}{\Lambda ^{4}}+y_{4}^{\left(
N\right) }\left( \overline{N}_{R}N_{R}^{C}\right) _{\overline{\mathbf{3}}%
_{S_{2}}}\phi \frac{\rho ^{4}}{\Lambda ^{4}}+y_{5}^{\left( N\right) }\left( 
\overline{N}_{R}N_{R}^{C}\right) \phi \frac{\rho ^{4}\eta ^{3}}{\Lambda ^{7}}%
+H.c. \,. \notag
\end{eqnarray}

\section{Lepton masses and mixings}

\label{leptonmixings} The spontaneous breaking of the electroweak gauge
symmetry as well as of the $\Delta (27)\times Z_{2}\times Z_{3}\times Z_{9}$
discrete group, gives rise to a diagonal mass matrix for charged leptons.
The resulting SM charged lepton masses take the form: 
\begin{equation}
m_{e}=y_{1}^{\left( l\right) }\frac{v_{\eta }v_{\sigma }^{8}v_{H_{d}}}{\sqrt{%
2}\Lambda ^{9}}=a_{1}^{\left( l\right) }\lambda ^{9}\frac{v}{\sqrt{2}},%
\hspace{1cm}m_{\mu }=y_{2}^{\left( l\right) }\frac{v_{\eta }v_{\sigma
}^{4}v_{H_{d}}}{\sqrt{2}\Lambda ^{5}}=a_{2}^{\left( l\right) }\lambda ^{5}%
\frac{v}{\sqrt{2}},\hspace{1cm}m_{\tau }=y_{3}^{\left( l\right) }\frac{%
v_{\eta }v_{\sigma }^{2}v_{H_{d}}}{\sqrt{2}\Lambda ^{3}}=a_{3}^{\left(
l\right) }\lambda ^{3}\frac{v}{\sqrt{2}} \,, \label{eq:lepmass}
\end{equation}%
with $a_{1}^{\left( l\right) }$, $a_{2}^{\left( l\right) }$ and $%
a_{3}^{\left( l\right) }$ being real $\mathcal{O}(1)$ dimensionless parameters and
we have assumed that $v_{H_{d}}\sim v/\sqrt{2}$, being $v=246$ GeV the
electroweak symmetry breaking scale.

Furthermore, from the neutrino Yukawa interactions we find the following
neutrino mass terms: 
\begin{equation}
-\mathcal{L}_{mass}^{\left( \nu \right) }=\frac{1}{2}\left( 
\begin{array}{ccc}
\overline{\nu _{L}^{C}} & \overline{\nu _{R}} & \overline{N_{R}}%
\end{array}%
\right) M_{\nu }\left( 
\begin{array}{c}
\nu _{L} \\ 
\nu _{R}^{C} \\ 
N_{R}^{C}%
\end{array}%
\right) +H.c.,
\end{equation}%
where the neutrino mass matrix is given by: 
\begin{equation}
M_{\nu }=\left( 
\begin{array}{ccc}
0_{3\times 3} & m_{\nu D} & 0_{3\times 3} \\ 
m_{\nu D}^{T} & 0_{3\times 3} & M \\ 
0_{3\times 3} & M^{T} & \mu%
\end{array}%
\right) ,  \label{Mnu}
\end{equation}%
and the submatrices $m_{\nu D}$, $M$ and $\mu $ have the following
structure: 
\begin{eqnarray}
m_{\nu D} &=&\frac{y_{\nu }v_{u}}{\sqrt{2}}\left( 
\begin{array}{ccc}
1 & 0 & 0 \\ 
0 & 1 & 0 \\ 
0 & 0 & 1%
\end{array}%
\right) ,\hspace{1cm}M=m\left( 
\begin{array}{ccc}
1 & 0 & 0 \\ 
0 & 1 & 0 \\ 
0 & 0 & 1%
\end{array}%
\right) +v_{\eta }\left( 
\begin{array}{ccc}
z_{1}+z_{2}+z_{3} & 0 & 0 \\ 
0 & z_{1}+\omega z_{2}+\omega ^{2}z_{3} & 0 \\ 
0 & 0 & z_{1}+\omega ^{2}z_{2}+\omega z_{3}%
\end{array}%
\right) \frac{v_{\eta }^{2}}{\Lambda ^{2}},\hspace{1cm}  \notag
\label{Mnublocks0} \\
\mu &=&y_{1}^{\left( N\right) }v_{\phi }\left( 
\begin{array}{ccc}
1 & 0 & 0 \\ 
0 & \omega & 0 \\ 
0 & 0 & \omega ^{2}%
\end{array}%
\right) \frac{v_{\sigma }^{5}}{\Lambda ^{5}}+y_{2}^{\left( N\right) }v_{\phi
}\left( 
\begin{array}{ccc}
0 & \omega ^{2} & \omega \\ 
\omega ^{2} & 0 & 1 \\ 
\omega & 1 & 0%
\end{array}%
\right) \frac{v_{\sigma }^{5}}{\Lambda ^{5}}  \notag \\
&&+y_{3}^{\left( N\right) }v_{\phi }\left( 
\begin{array}{ccc}
1 & 0 & 0 \\ 
0 & \omega ^{2} & 0 \\ 
0 & 0 & \omega%
\end{array}%
\right) \frac{v_{\rho }^{4}}{\Lambda ^{4}}+y_{4}^{\left( N\right) }v_{\phi
}\left( 
\begin{array}{ccc}
0 & \omega & \omega ^{2} \\ 
\omega & 0 & 1 \\ 
\omega ^{2} & 1 & 0%
\end{array}%
\right) \frac{v_{\rho }^{4}}{\Lambda ^{4}} \,.
\end{eqnarray}%
The tiny masses of the light active neutrinos are due to the inverse
seesaw mechanism, which yields the following mass matrices for the physical
neutrino fields: 
\begin{eqnarray}
\widetilde{M}_{\nu } &\simeq &m_{\nu D}\left( M^{T}\right) ^{-1}\mu
M^{-1}m_{\nu D}^{T}\simeq \left( 
\begin{array}{ccc}
A_{1}+A_{2} & B_{1}\omega ^{2}+B_{2}\omega & B_{1}\omega +B_{2}\omega ^{2}
\\ 
B_{1}\omega ^{2}+B_{2}\omega & A_{1}\omega +A_{2}\omega ^{2} & B_{1}+B_{2}
\\ 
B_{1}\omega +B_{2}\omega ^{2} & B_{1}+B_{2} & A_{1}\omega ^{2}+A_{2}\omega%
\end{array}%
\right) ,\hspace{0.7cm}  \label{M1nu} \\
M_{\nu }^{\left( -\right) } &=&-\frac{1}{2}\left( M+M^{T}\right) +\frac{1}{2}%
\mu ,\hspace{0.7cm} \\
M_{\nu }^{\left( +\right) } &=&\frac{1}{2}\left( M+M^{T}\right) +\frac{1}{2}%
\mu \,,  \label{neutrino-mass}
\end{eqnarray}%
where $\widetilde{M}_{\nu }$ corresponds to the mass matrix for the light active neutrinos, whereas $M_{\nu }^{(-)}$ and $M_{\nu }^{(+)}$ are the mass
matrices for the heavy neutrinos. The physical neutrino spectrum includes 3
light active neutrinos as well as 6 nearly degenerate sterile exotic
pseudo-Dirac neutrinos. The limit $\mu \rightarrow 0$ restores lepton number
symmetry, yielding massless active neutrinos. In this regime, the sterile
neutrinos experience a suppressed mass splitting and thus behave as
pseudo-Dirac leptons.

The physical observables of the low energy neutrino sector, i.e., the two
neutrino mass squared differences, the three leptonic mixing parameters and
the leptonic Dirac CP violating phase can be very well reproduced in our
model. For the scenario of normal and inverse neutrino mass hierarchy we obtain the
following best-fit point shown in Table \ref{tab:neutrino_fit}: 

\begin{table}[H]
\centering
\begin{tabular}{l||c|c|c}
 Parameter & Global Fit at $\pm1\sigma$ NO (IO)  & Model Value NO  & Model Value IO \\ \hline \hline
 $\Delta m^2_{21}\ [\times10^{-5}\mathrm{eV}^2]$ & $7.55^{+0.22}_{-0.20}$ & $7.31$ & $7.81$ \\ \hline
 $\Delta m^2_{31}\ [\times10^{-5}\mathrm{eV}^2]$ & $2.50\pm0.02\ \left(2.40\pm0.02 \right)$ & $2.46$ & $2.38$ \\ \hline
 $\sin^2 \theta_{12} \times 10^{-1}$ & $3.04\pm0.16$ & $2.87$ & $3.37$  \\ \hline
 $\sin^2 \theta_{23} \times 10^{-1}$& $5.60^{+0.13}_{-0.22}\ \left(5.57^{+0.14}_{-0.20} \right) $ & $4.78$ & $5.90$ \\ \hline
 $\sin^2 \theta_{13}\times 10^{-2}$& $2.20^{+0.07}_{-0.04}\ \left( 2.23^{+0.05}_{-0.06} \right)$ & $2.28$ & $2.28$  \\ \hline
 $\delta_{CP}\ [\mathrm{rad}]$ & $1.12^{+0.16}_{-0.12}\ \left( 1.50^{+0.13}_{-0.14} \right)$ & $0.851$ & $1.39$
\end{tabular}
\caption{Comparison between Global Fit at $\pm 1\sigma$ \cite{Tortola:2024} and the Model Values in normal ordering (NO) and inverse ordering (IO). All the values are in $3\sigma$ range.}
\label{tab:neutrino_fit}
\end{table}
In addition to the Global Fit, the atmospheric mixing parameter $\sin^2\theta_{12}$ predicted by our model is fully compatible at $3\sigma$ with the recent results of the JUNO experiment \cite{JUNO:2025gmd}, corresponding to 
the value of $\sin^2\theta_{12}=0.3092\pm 0.0087$.
To successfully reproduce these values, the effective parameters of the low energy neutrino neutrino mass matrix for normal and inverse orderings are:
\begin{align}
\mathrm{\textbf{NO}} :& \
A_{1} \approx 1.28\times 10^{-2}\left[ eV\right] ,\ \ A_{2}\approx
1.78\times 10^{-3}\left[ eV\right] ,\ \ B_{1}\approx \left( -1.46\times
10^{-2}+-7.42\times 10^{-3}i\right) \left[ eV\right] ,  \notag \\
& \ B_{2} \approx \left( -7.35\times 10^{-3}-1.68\times 10^{-3}i\right) [eV], 
\notag \\
\mathrm{\textbf{IO}} :& \ A_{1} \approx 2.74\times 10^{-2}\left[ eV\right] ,\ \ \ \ A_{2}\approx
5.65\times 10^{-3}\left[ eV\right] ,\ \ \ \ \ B_{1}\approx \left(
-2.18\times 10^{-2}+-10.1\times 10^{-2}i\right) \left[ eV\right] ,\ \  
\notag \\
& \ B_{2} \approx \left( -9.12\times 10^{-3}-1.49\times 10^{-2}i\right) \left[
eV\right]
\label{eq:benchmark}
\end{align}%
As shown in Eq. (\ref{eq:benchmark}), we consider $A_1$, $A_2$ to be real and $B_1$, $B_2$ to be complex. To find these values, we scan in the range $[-1,1]$ [eV] for the real and imaginary parts that fit the neutrino observables.
Our model also predicts the effective Majorana mass $m_{ee}$, which governs
neutrinoless double beta decay ($0\nu \beta \beta $) and tests the Majorana
nature of neutrinos. It is defined as 
\begin{equation}
m_{ee}=\left\vert\sum_{i=1}^3\mathbf{U}_{ei}^{2}\,m_{\nu_i}\right\vert \,,
\label{ec:mee}
\end{equation}%
where $\mathbf{U}_{ei}$ are PMNS mixing elements and $m_{\nu_i}$ ($i=1,2.3$)
the light active neutrino masses. Furthermore, we also find predictions for
the parameter $m_{e}$ defined as: 
\begin{equation}
m_{e}=\sqrt{\sum_{i=1}^3\left\vert\mathbf{U}_{ei}\right\vert ^{2}\,m_{\nu_i}^{2}%
} \,. \label{ec:mee}
\end{equation}%
\begin{figure}[tbp]
\includegraphics[width=\textwidth]{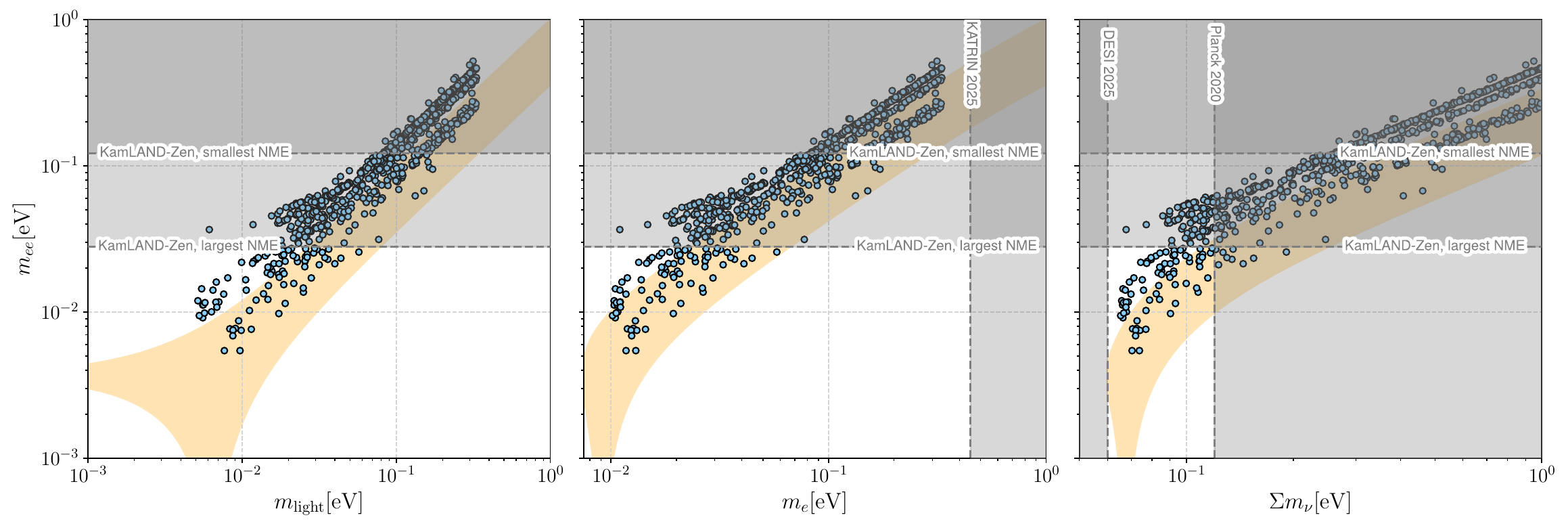}
\caption{Correlations of the effective Majorana mass parameter of neutrinoless double beta decay $m_{ee}$ with the lightest neutrino mass, the parameter $m_{e}$ and the sum of the active neutrino masses for the scenario of normal neutrino mass hierarchy. The limits in dashed line correspond to KamLAND-Zen \cite{KamLAND-Zen:2024eml}, KATRIN2025 \cite{Schlosser:2025vla}, DESI2025 \cite{Jiang:2024viw} and Planck2020 \cite{Planck:2018vyg}.}
\label{meeNH}
\end{figure}
\begin{figure}[tbp]
\includegraphics[width=\textwidth]{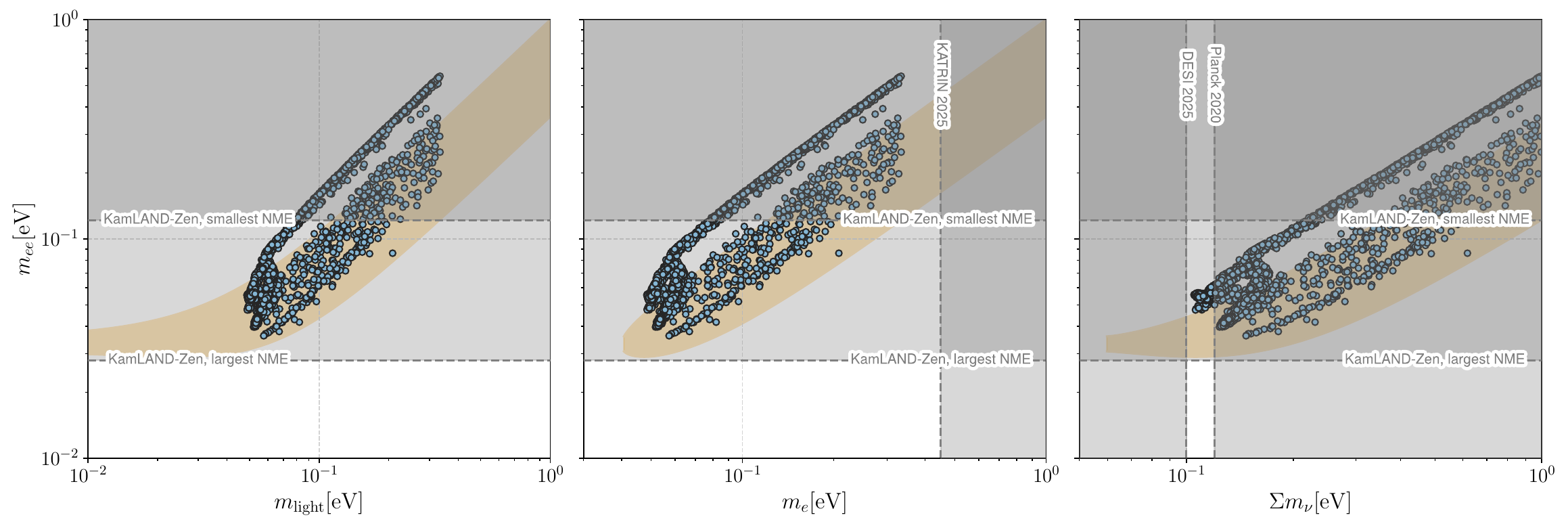}
\caption{Correlations of the effective Majorana mass parameter of neutrinoless double beta decay $m_{ee}$ with the lightest neutrino mass, the parameter $m_{e}$ and the sum of the active neutrino masses for the scenario of inverted neutrino mass hierarchy. The limits in dashed line correspond to KamLAND-Zen \cite{KamLAND-Zen:2024eml}, KATRIN2025 \cite{Schlosser:2025vla}, DESI2025 \cite{Jiang:2024viw} and Planck2020 \cite{Planck:2018vyg}.}
\label{meeIH}
\end{figure}
Figures~\ref{meeNH} and \ref{meeIH} display the correlations of the effective Majorana mass parameter of neutrinoless double beta decay $m_{ee}$ with the lightest neutrino mass, the parameter $m_{e}$ and the sum of the active neutrino masses $\sum_{i=1}^3 m_{\nu_i}$, for the scenarios of normal and inverted neutrino mass hierarchies, respectively. In generating these Figures, we used the model parameters consistent with the $3\sigma $ experimentally allowed ranges. For
normal (NO) and inverted (IO) neutrino mass orderings, the predicted ranges are respectively given by: 
\begin{align}
4.1\ \text{meV}&\lesssim m_{ee}\lesssim 5.4\times 10^{2} \ \text{meV},\hspace{1cm}\text{for NO},
\notag \\
3.6\times 10^{1} \ \text{meV} &\lesssim m_{ee}\lesssim 5.5\times 10^{2}\ \text{meV} ,\hspace{1cm}\text{for IO} \,.
\end{align}%
The current strongest limit, $m_{ee}\leq \left( 28 - 122\right)\ \text{meV}$, comes from
KamLAND-Zen half-life bound $T_{1/2}^{0\nu \beta \beta }(^{136}\text{Xe}%
)>3.8\times 10^{26}$ yr~\cite{KamLAND-Zen:2024eml}. These limits are shown in Figure \ref{meeNH} and \ref{meeIH}, which shows that the model is consistent with these limits only in the NO scenario.

Regarding the sum of active neutrino masses, the model yield the following predictions: 
\begin{eqnarray}
&&0.06\ \text{eV}\lesssim \sum_{i=1}^3 m_{\nu_i}\lesssim 0.99\ \text{eV},\hspace{1cm}\text{for NO%
},  \notag \\
&&0.10\ \text{eV}\lesssim \sum_{i=1}^3 m_{\nu_i}\lesssim 0.99\ \text{eV},\hspace{1cm}\text{for IO%
} \,,
\end{eqnarray}%
with a best-fit values $\sum_{i=1}^3 m_{\nu_i}\simeq 0.066\ \text{eV}$ and $\sum_{i=1}^3 m_{\nu_i}\simeq 0.11 \ 
\text{eV}$ for the scenarios of normal and inverted neutrino mass hierarchies,
respectively. This lies within recent cosmological bounds of roughly $%
0.04-0.3$ eV~\cite{Jiang:2024viw,Naredo-Tuero:2024sgf}.

\section{Leptogenesis}

\label{lepto} In this section, we examine the consequences of our model for
leptogenesis. For simplicity, we consider the contribution to the lepton
asymmetry arising from the lightest pseudo-Dirac fermions $N_1^\pm \equiv
N^\pm$. The CP-violating lepton asymmetry generated by the decays of $N^\pm$ is given by \cite{Gu:2010xc,Pilaftsis:1997jf}:  
\begin{equation}
\varepsilon_\pm \equiv \sum_{i=1}^{3} \frac{ \Gamma\!\left(N_\pm \to \nu_i
h\right) - \Gamma\!\left(N_\pm \to \overline{\nu}_i h\right) }{
\Gamma\!\left(N_\pm \to \nu_i h\right) + \Gamma\!\left(N_\pm \to \overline{%
\nu}_i h\right) } \simeq \frac{ \text{Im}\!\left\{ \left[ \left(y_{N_+}%
\right)^\dagger \left(y_{N_-}\right) \right]^2_{11} \right\} }{8\pi A_\pm} 
\frac{r}{r^2+\Gamma_\pm^2/m_{N_\pm}^2} \,,
\end{equation}
where 
\begin{align}
r &= \frac{m_{N_+}^2 - m_{N_-}^2}{m_{N_+} m_{N_-}}, & A_\pm &= \left[%
\left(y_{N_\pm}\right)^\dagger y_{N_\pm}\right]_{11}, \\
y_{N_\pm} &= y_\nu (1 \mp S) = y_\nu \!\left(1 \pm \frac{1}{4} M^{-1}\mu
\right), & \Gamma_\pm &= \frac{A_\pm m_{N_\pm}}{8\pi} \,.
\end{align}
If interference effects between the two sterile states $N^\pm$ are
neglected, the total washout parameter $K_{N^+}+K_{N^-}$ becomes very large,
as pointed out in Ref.~\cite{Dolan:2018qpy}. However, the small mass
splitting characteristic of pseudo-Dirac neutrinos induces destructive
interference in the relevant scattering processes~\cite{Blanchet:2009kk}.
Accounting for this effect, the effective washout parameter can be written
as 
\begin{equation}
K^{\text{eff}} \simeq K_{N^+}\,\delta_+^2 + K_{N^-}\,\delta_-^2 \,,
\end{equation}
with 
\begin{equation}
\delta_\pm = \frac{m_{N^+}-m_{N^-}}{\Gamma_{N^\pm}}, \qquad K_{N^\pm} = 
\frac{\Gamma_\pm}{H(m_{N_\pm})} \,.
\end{equation}
Here $H$ denotes the Hubble expansion rate.

Assuming a standard cosmological history in which the energy density is
dominated by SM radiation, the Hubble parameter is given by 
\begin{equation}  \label{eq:Hubble}
H(T) = \frac{\pi}{3}\sqrt{\frac{g_\star}{10}}\frac{T^2}{M_P} \,,
\end{equation}
where $g_\star$ is the effective number of relativistic degrees of freedom
in the thermal bath, and $M_P \simeq 2.4 \times 10^{18}\,\text{GeV}$ is the
reduced Planck mass.

\begin{figure}[h]
\centering
\begin{minipage}{0.48\textwidth}
		\centering
		\includegraphics[width=\linewidth]{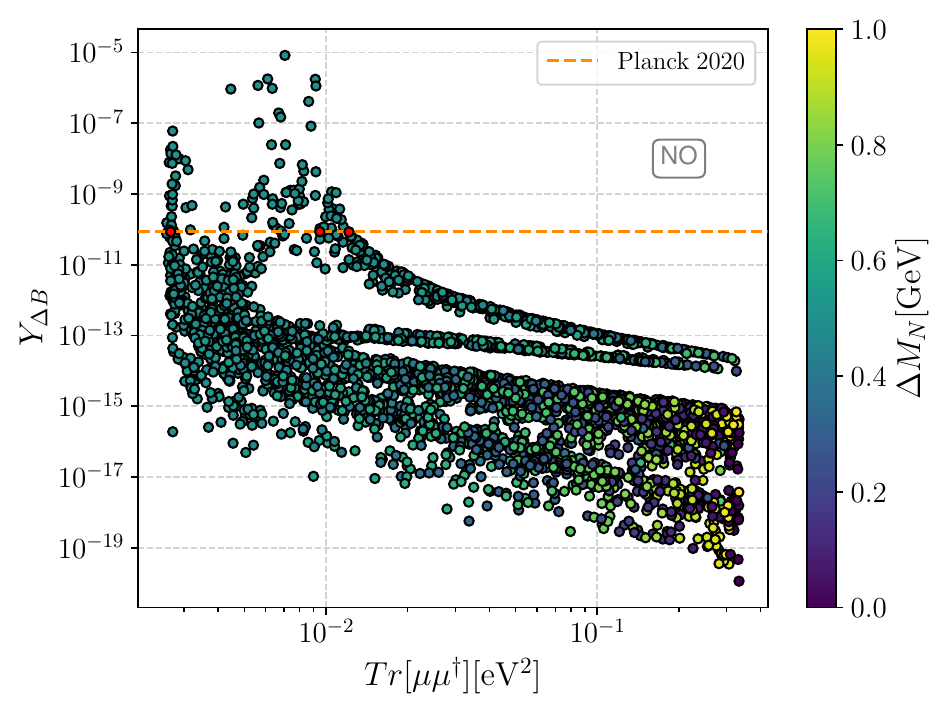}
		\par\vspace{2pt}
			\end{minipage} \hfill 
\begin{minipage}{0.48\textwidth}
		\centering
		\includegraphics[width=\linewidth]{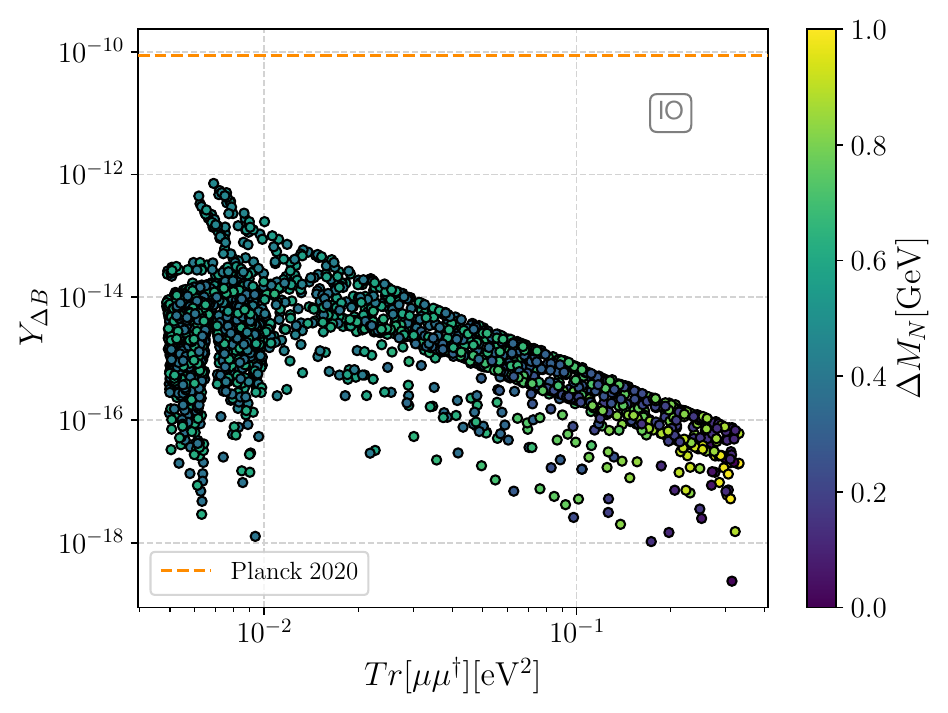}
		\par\vspace{2pt}
			\end{minipage}
\caption{The baryon asymmetry parameter as a function of $Tr\left[ \protect%
\mu \protect\mu ^{\dagger }\right] $ for normal (left panel) and inverted
(right panel) neutrino mass hierarchy. The red points in left panel are the points that fit the baryon asymmetry parameter $Y_{\Delta B}=(0.87\pm 0.01)\times 10^{-10}$.}
\label{YB}
\end{figure}

In both the weak- and strong-washout regimes, the resulting baryon asymmetry
is related to the total lepton asymmetry as~\cite{Pilaftsis:1997jf} 
\begin{align}
Y_{\Delta B}& \equiv \frac{n_{B}-\overline{n}_{B}}{s}=-\frac{28}{79}\frac{%
\epsilon _{+}+\epsilon _{-}}{g^{\ast }},\qquad & & K^{\text{eff}}\ll 1, \\
Y_{\Delta B}& \equiv \frac{n_{B}-\overline{n}_{B}}{s}=-\frac{28}{79}\frac{%
0.3(\epsilon _{+}+\epsilon _{-})}{g^{\ast }K^{\text{eff}}(\ln K^{\text{eff}%
})^{0.6}},\qquad & & K^{\text{eff}}\gg 1 \,.
\end{align}%
In our numerical analysis, we impose the requirement that our model
successfully accommodates the measured value of the baryon asymmetry of the
Universe measured by Planck~\cite{Planck:2018vyg}, 
\begin{equation}
Y_{\Delta B}=(0.87\pm 0.01)\times 10^{-10}.
\end{equation}%
Our predictions for the baryon asymmetry parameter $Y_{\Delta B}$ as a
function of $Tr\left[ \mu \mu ^{\dagger }\right] $ for normal and inverted
neutrino mass hierarchies are displayed in the left and right panels of
Figure \ref{YB}, respectively. As indicated in Figure \ref{YB} our model is
compatible with the experimental range of the baryon asymmetry parameter in the NO scenario, whereas the IO is discarded. Figure \ref{YB}
provided that the lepton number violating Majorana submatrix $\mu $ fulfils
the constraint  $
10^{-3}eV^2\lesssim Tr\left[ \mu \mu ^{\dagger }\right] \lesssim 10^{-1}eV^2$. 

\section{Conclusions}

\label{conclusions}

We have proposed a supersymmetric low scale seesaw model where the tiny
masses of the active neutrinos arise from an inverse seesaw model and the
leptonic mixing is governed by the cobimaximal pattern. This is achieved by
extending the SM gauge symmetry by the inclusion of the $\Delta (27)\times
Z_{2}\times Z_{3}\times Z_{9}$ discrete group. Additionally, achieving the
cobimaximal mixing pattern requires the inclusion of six Majorana neutral
leptons and a moderate amount of gauge singlet scalars. Besides that, in our
proposed model, the SM charged lepton mass hierarchy is caused by the
spontaneous breaking of the $Z_{9}$ discrete group. We have shown that our
model is compatible with the experimental values of neutrino mass squared splittings, leptonic mixing parameters and leptonic Dirac CP violating phase 
for both scenarios of normal and inverted neutrino mass hierarchies. Besides that, we have found that our model successfully complies with neutrinoless double beta decay constraints only for the scenario of normal neutrino mass hierarchy. Furthermore, we have shown that our model is capable of successfully reproducing the measured value of the baryon asymmetry observed in the
Universe only in the scenario of normal neutrino mass ordering provided that
the submatrix $\mu $ associated with the lepton number violation satisfies $%
10^{-3}eV^2\lesssim Tr\left[ \mu \mu ^{\dagger }\right] \lesssim 10^{-1}eV^2$. 

\section{Acknowledgments}

The authors are very grateful to Vishnudath Neelakandan for his involvement in the initial stages of the project and for very useful discussions. AECH is supported by ANID-Chile FONDECYT 1261103, 1241855, ANID Millennium Science
Initiative Program $ICN2019\_044$, ANID CCTVal CIA250027 and ICTP through
the Associates Programme (2026-2031).
AECH thanks the Instituto
Superior T\'{e}cnico, Universidade de Lisboa and ICTP for hospitality, where
part of this work was done. %
NAPJ is supported by ANID-Chile Doctorado Nacional año 2022 21221396, and Programa de Incentivo a la Investigaci\'on Cient\'{\i}fica (PIIC) from UTFSM.
IdMV thanks the University of Basel for hospitality.
IdMV acknowledges funding from Fundação para a Ciência e a Tecnologia (FCT) through the FCT Mobility program, and through
the projects CFTP-FCT Unit 
UID/00777/2025 (\url{https://doi.org/10.54499/UID/00777/2025}),  UIDB/FIS/00777/2020 and UIDP/FIS/00777/2020, CERN/FIS-PAR/0019/2021,
CERN/FIS-PAR/0002/2021, 2024.02004 CERN, which are partially funded through POCTI (FEDER), COMPETE,
QREN and EU.

\appendix

\section{The $\Delta (27)$ discrete group}

\label{delta27} The $\Delta (27)$ discrete group has the following 11
irreducible representations: one triplet $\mathbf{3}$, one antitriplet $%
\overline{\mathbf{3}}$ and nine singlets $\mathbf{1}_{k,l}$ ($k,l=0,1,2$),
where $k$ and $l$ identify how the singlets transform under order 3
generators, corresponding to a $Z_{3}$ and $Z_{3}^{\prime }$ subgroups of $%
\Delta (27)$. 
\begin{eqnarray}
\mathbf{3}\otimes \mathbf{3} &=&\overline{\mathbf{3}}_{S_{1}}\oplus 
\overline{\mathbf{3}}_{S_{2}}\oplus \overline{\mathbf{3}}_{A}  \notag \\
\overline{\mathbf{3}}\otimes \overline{\mathbf{3}} &=&\mathbf{3}%
_{S_{1}}\otimes \mathbf{3}_{S_{2}}\oplus \mathbf{3}_{A}  \notag \\
\mathbf{3}\otimes \overline{\mathbf{3}} &=&\sum_{r=0}^{2}\mathbf{1}%
_{r,0}\oplus \sum_{r=0}^{2}\mathbf{1}_{r,1}\oplus \sum_{r=0}^{2}\mathbf{1}%
_{r,2}  \notag \\
\mathbf{1}_{k,\ell }\otimes \mathbf{1}_{k^{\prime },\ell ^{\prime }} &=&%
\mathbf{1}_{k+k^{\prime }mod3,\ell +\ell ^{\prime }mod3} \\
&&
\end{eqnarray}%
Denoting $\left( x_{1},y_{1},z_{1}\right) $ and $\left(
x_{2},y_{2},z_{2}\right) $ as the basis vectors for two $\Delta (27)$%
-triplets $\mathbf{3}$, one finds: 
\begin{eqnarray}
\left( \mathbf{3}\otimes \mathbf{3}\right) _{\overline{\mathbf{3}}_{S_{1}}}
&=&\left( x_{1}y_{1},x_{2}y_{2},x_{3}y_{3}\right) ,  \notag
\label{triplet-vectors} \\
\left( \mathbf{3}\otimes \mathbf{3}\right) _{\overline{\mathbf{3}}_{S_{2}}}
&=&\frac{1}{2}\left(
x_{2}y_{3}+x_{3}y_{2},x_{3}y_{1}+x_{1}y_{3},x_{1}y_{2}+x_{2}y_{1}\right) , 
\notag \\
\left( \mathbf{3}\otimes \mathbf{3}\right) _{\overline{\mathbf{3}}_{A}} &=&%
\frac{1}{2}\left(
x_{2}y_{3}-x_{3}y_{2},x_{3}y_{1}-x_{1}y_{3},x_{1}y_{2}-x_{2}y_{1}\right) , 
\notag \\
\left( \mathbf{3}\otimes \overline{\mathbf{3}}\right) _{\mathbf{1}_{r,0}}
&=&x_{1}y_{1}+\omega ^{2r}x_{2}y_{2}+\omega ^{r}x_{3}y_{3},  \notag \\
\left( \mathbf{3}\otimes \overline{\mathbf{3}}\right) _{\mathbf{1}_{r,1}}
&=&x_{1}y_{2}+\omega ^{2r}x_{2}y_{3}+\omega ^{r}x_{3}y_{1},  \notag \\
\left( \mathbf{3}\otimes \overline{\mathbf{3}}\right) _{\mathbf{1}_{r,2}}
&=&x_{1}y_{3}+\omega ^{2r}x_{2}y_{1}+\omega ^{r}x_{3}y_{2},
\end{eqnarray}%
where $r=0,1,2$ and $\omega =e^{i\frac{2\pi }{3}}$.

\appendix

\bibliographystyle{utphys}
\bibliography{RefsISDelta27.bib}

\end{document}